# Conceptual Modeling for Control of a Physical Engineering Plant: A Case Study

Sabah Al-Fedaghi

Computer Engineering Department
Kuwait University
Kuwait
sabah.alfedaghi@ku.edu.kw

Abdulaziz AlQallaf

Instrument Maintenance Department
Ministry of Electricity and Water
Kuwait
Alqallaf.AQ@gmail.com

*Abstract*—We examine the problem of weaknesses in frameworks of conceptual modeling for handling certain aspects of the system being modeled. We propose the use of a flow-based modeling methodology at the conceptual level. Specifically, and without loss of generality, we develop a conceptual description that can be used for controlling the maintenance of a physical system, and demonstrate it by applying it to an existing electrical power plant system. Recent studies reveal difficulties in finding comprehensive answers for monitoring operations and identifying risks as well as the fact that incomplete information can easily lead to incorrect maintenance. A unified framework for integrated conceptualization is therefore needed. The conceptual modeling approach integrates maintenance operations into a total system comprising humans, physical objects, and information. The proposed model is constructed of (abstract) machines of "things" connected by flows, forming an integrated whole. It represents a man-made, intentionally constructed system and includes technical and human "things" observable in the real world, exemplified by the study case described in this paper. A specification is constructed from a maximum of five basic operations: creation, processing, releasing, transferring, and receiving.

## I. INTRODUCTION

The use of models is an important aspect of engineering disciplines because of their essential role in understanding and engaging with the world [1]. Models can take many forms:

We can use words, drawings or sketches, physical models, computer programs, or mathematical formulas. In other words, the modeling activity can be done in several languages, often simultaneously. [2]

Accordingly, models can be classified as different types: conceptual, physical, or mathematical [3]. In this paper, we focus on conceptual models used to capture "conceptual structures of a domain" [4].

"A model is an abstract view of portion of reality that assists developers to concentrate on relevant aspects of the system and discount needless complications" [5]. ISO/IEC/IEEE 42010 (2011) [6] defines a model as follows:

"M is a model of S if M can be used to answer questions about S." In principle, a model is anything that can describe a system, and in this sense, all kinds of typical engineering work products that are created to specify or describe a system are models [7]. The major advantage of modeling is that models are expressed in terms of concepts bound much less to the underlying implementation technology and more closely to the problem domain [1].

ISO/IEC/IEEE 15288 (2015) [8] defines a *system* as a combination of interacting system elements organized to achieve one or more stated purposes. In this paper, we view a system as an (abstract) machine of "things" (to be defined later) connected by flows to form an integrated whole. The machine represents a man-made, intentionally constructed system (hence, it has a purpose) and includes technical and human "things" observable in the real world, as we exemplify in a case study in this paper. "Things" can be pipes, valves, structures, events and happenings, procedures, or materials, e.g., water, chlorine, and heat. A machine is constructed from at most *five* basic operations: creation, processing, releasing, transferring, and receiving. In this paper, we focus on the control and tracking of flows of "things" through machines for maintenance, operations, and management.

Conceptual modeling is a phase of system development that usually occurs after requirements analysis and precedes the design phase in the life cycle of "things". The conceptual model is constituted of a structure that reflects the composition of the physical elements of the system, and behavior that specifies the operational scenarios and functions of the system [7]. It facilitates understanding and communication among stakeholders and serves as a base for consequent phases. Valued features in conceptual models include completeness, faithfulness to realization of the system, understandability, and susceptibility analysis.

Most current conceptual modeling techniques use object-oriented methodology (e.g., UML, SysML), because their main foundation requires breaking system behavior into several pieces and then further decomposing those into other diagrams. Many claims have been made regarding the benefits of an object-oriented model, such as simulating the modeler's way of thinking [9] and contributing to "reducing complexity in the representation of technical systems and design processes" [10]. Researchers have examined and proposed extending the use of object-oriented languages such as UML, but Evermann [11] notes that "UML is suitable for conceptual modeling but the modeler must take special care not to confuse software aspects with aspects of the real world being modeled." The problem with extending UML is that "[UML] possesses no real-world business or organizational meaning; i.e., it is unclear what the constructs of such languages mean in terms of the business" [11]. The object-oriented modeling domain deals with objects and attributes, whereas the real-world domain deals with things and properties. According to Mordecai [12], there is a "significant inability of common conceptual modeling frameworks to appeal to practicing designers and analysts." These frameworks have an "inherent limitation and even fixation to handling the nominal view of the system being modeled. ... A unified framework for integrated, multipurpose, robust, and disruption-accommodating modeling and management is therefore urgently needed" [12].

In contrast to the object-oriented paradigm, according to Dori [13], models of complex systems should conveniently combine *structure* and *behavior* in a single model. Object-Process Methodology (OPM) [13] was developed for multidisciplinary, complex, and dynamic systems and processes [12]. OPM is chartered as ISO/PAS 19450 for system and process modeling [14]. It is considered "a state-of-the-art methodology and paradigm" in both the conceptual modeling domain [15] and the model-based systems engineering domain [16]. OPM [13] is a holistic approach to modeling, studying, and developing engineering systems.

The OPM paradigm integrates the object-oriented, process-oriented, and state transition approaches into a single frame of reference. *Structure and behavior coexist in the same OPM model without highlighting one at the expense of suppressing the other* to enhance the comprehension of the system as a whole. [17] (Italics added)

In this paper, we introduce an alternative to object-oriented and object-process methodologies, a conceptual modeling methodology based on *flows*, and also present a different conceptualization of such notions as processes, things (objects), and events. To show the viability of the proposed methodology, and without loss of generality, we develop a conceptual description that can be used for control of maintenance and operations of a physical system; as an example, we use the flow of operations within an existing electrical power station. *Maintenance* here refers to "actions taken to prevent a system structure or component from failing or to repair normal equipment degradation experienced with the operation of the device to keep it in proper working order" [18]. *Operations* ensure [19] the implementation and control of activities and safe and reliable processes, as well as recognition of the status of all equipment and operators' knowledge and performance; this aspect supports safe and reliable plant operation.

Recent studies reveal difficulties in finding comprehensive answers to problems inherent in monitoring of operations in physical systems, such as identifying risks and difficulties related to incomplete data, which can lead to incorrect maintenance and operations [20-21]. According to

Vieira and Marques [22], "The definition of policies and strategies and the understanding of the efficiency and effectiveness of the maintenance department continue to present opportunities for improvement" [22].

A unified conceptual framework (a single diagram) seems to provide many benefits, e.g., completeness, understandability, and simplified analysis, and is a potential solution to the problems mentioned in the previous paragraph. The conceptual modeling approach integrates maintenance and operations into a total system that comprises humans, physical objects, and information. In our study case, as a result of the complexity of maintenance management of the electrical power plant, operations such as maintenance and technical management are no longer considered mere technical matters. Hence, operations must be integrated into the total management of the system, and a system for future possible online control must be developed. To this end, a conceptual description of the site is needed to provide a holistic overview of the various processes in the system.

## II. FLOWTHING MACHINE

For the sake of a self-contained paper, in this section, in subsection A, we briefly review our proposed methodology, which forms the foundation of the theoretical development in this paper called the Flowthing Machine (FM). It involves a diagrammatic language that has been adopted in several applications [23-31]. In subsection B, we provide a new example to explain the approach more completely.

### A. Basic Model

The FM modeling language is a uniform method for representing "things" that flow, called "flow things". Flow in the FM refers to the exclusive (i.e., being in one and only one) transformation among five states (also called stages): *transfer, process, create, release*, and *receive*. A flow thing (hereafter a *thing*) cannot be in two stages simultaneously. A thing is defined as what is created, released, transferred, received, and processed. Things in stages are analogous to molecules of water being in one of three states while in Earth's atmosphere: solid, liquid, or gas.

Each *stage* can be expressed by many words:

- *Create:* generate, appear (in the system), produce, make . . .
- *Transfer:* transport, communicate, send, transmit ...
- *Process:* millions of English verbs that change the form of a thing without creating a new one, e.g., paint, package, categorize . . .

Notions in FM can be described as follows.

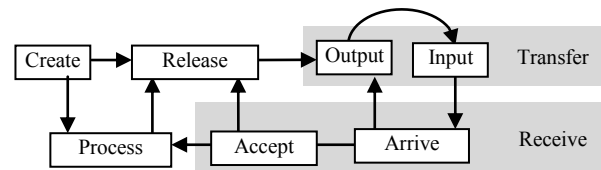

Fig. 1. Flow machine.

A flow machine (hereafter *machine*) is depicted in Fig. 1, which shows the internal flow of a system along with the five stages and transactions among them. The machine displayed in Fig. 1 is a generalization of the typical input-process-output model used in many scientific fields.

- *Spheres and subspheres:* These are the network environments and relationships of machines and submachines. The FM model represents a web of interrelated *flows* that cross the boundaries of intersecting and nested *spheres*. A particular static model is the space context for *happenings*, as will be explained later.

- *Triggering:* Triggering is a transformation (denoted by a dashed arrow) from one flow to another; e.g., a flow of electricity triggers a flow of air.

### B. Example

Rahim et al. [32] proposed a transformation to derive a modular Petri net from SysML activities to formalize and verify SysML requirements. They present a case study of the operation of a ticket vending machine (TVM):

The behaviour of the machine is triggered by passengers who need to buy a ticket. When a passenger starts a session, the TVM will request trip information from commuter. Passengers use the front panel to specify their boarding and destination place, details of passengers (number of adults and children) and date of travel. Based on the provided trip info, the TVM will calculate payment due and display the fare for the requested ticket. Then, it requests payment options. Those options include payment by cash, or by credit or debit card. After that, the passenger chooses a payment option and processes to payment. After a successful payment, the TVM prints and provides a ticket to the passenger.

Of interest in the present paper is the type of diagram used by Rahim et al. [32]. The activity diagram utilizes a composite activity concept that incorporates other activities, as shown partially in Fig. 2.

The purpose of scrutinizing this figure is not to present a fair description of Rahim et al.'s [32] study; rather, the aim is to visually contrast their activity diagrams with FM diagrams without a detailed comparison to demonstrate that the latter can be appreciated for their simple visual appearance and understandability.

### C. Static Description

Fig. 3 shows the FM representation of TVM activities. It comprises two main spheres: the passenger (number 1 in the figure) and the TVM (2). The passenger creates a request to start (3) that flows to the machine (4), where it is processed (5) to trigger (6) the generation of a message to input information (7). The message flows to the customer (8) to be processed to create the requested information (9-10), which then flows to the TVM (11). There, the information is processed (12) to trigger a payment transaction that includes

- creating the amount of payment (13) and
- creating the selection of payment options (14).

The payment amount and options flow to the traveler (15) to be processed (16) and to trigger selection of a payment method (17).

The selection flows to the TVM (18), where it is processed (19); depending on the type of payment,

- if the selection is for a cash payment, this triggers (20) the creation of a message to insert cash that flows to the passenger (21) to trigger the passenger to "create" (22; produce) cash that flows to the TVM (23). The cash is processed (24) as follows:

(a) If the cash is not sufficient, then the TVM creates a message (25) to complete the amount and sends it to the passenger (26).

(b) If the cash is correct, then the TVM triggers the creation (27) of tickets and sends them to the passenger (28).

(c) If the passenger decides to cancel the transaction and generates a signal (29) to refund the cash, then the TVM releases (30) the cash back to the passenger.

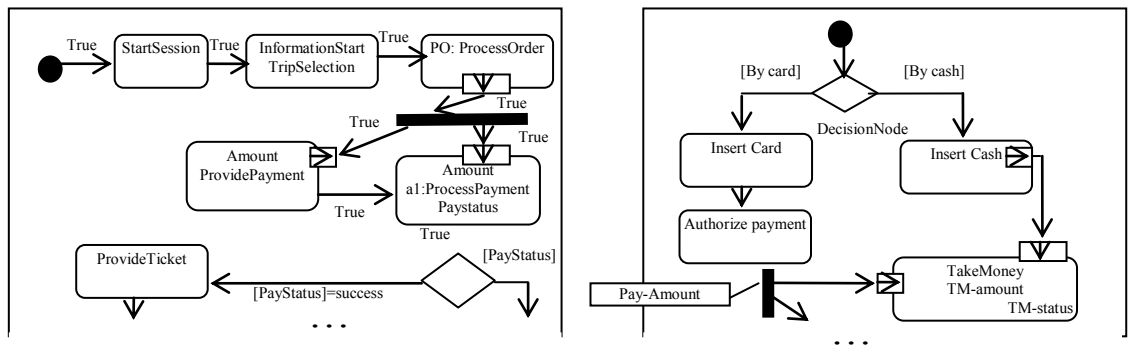

(a)   A main activity diagram for TVM.    (b) A subactivity diagram of the payment process.

Fig. 2. Main and subactivity diagrams for TVM (redrawn, partial from [32]).

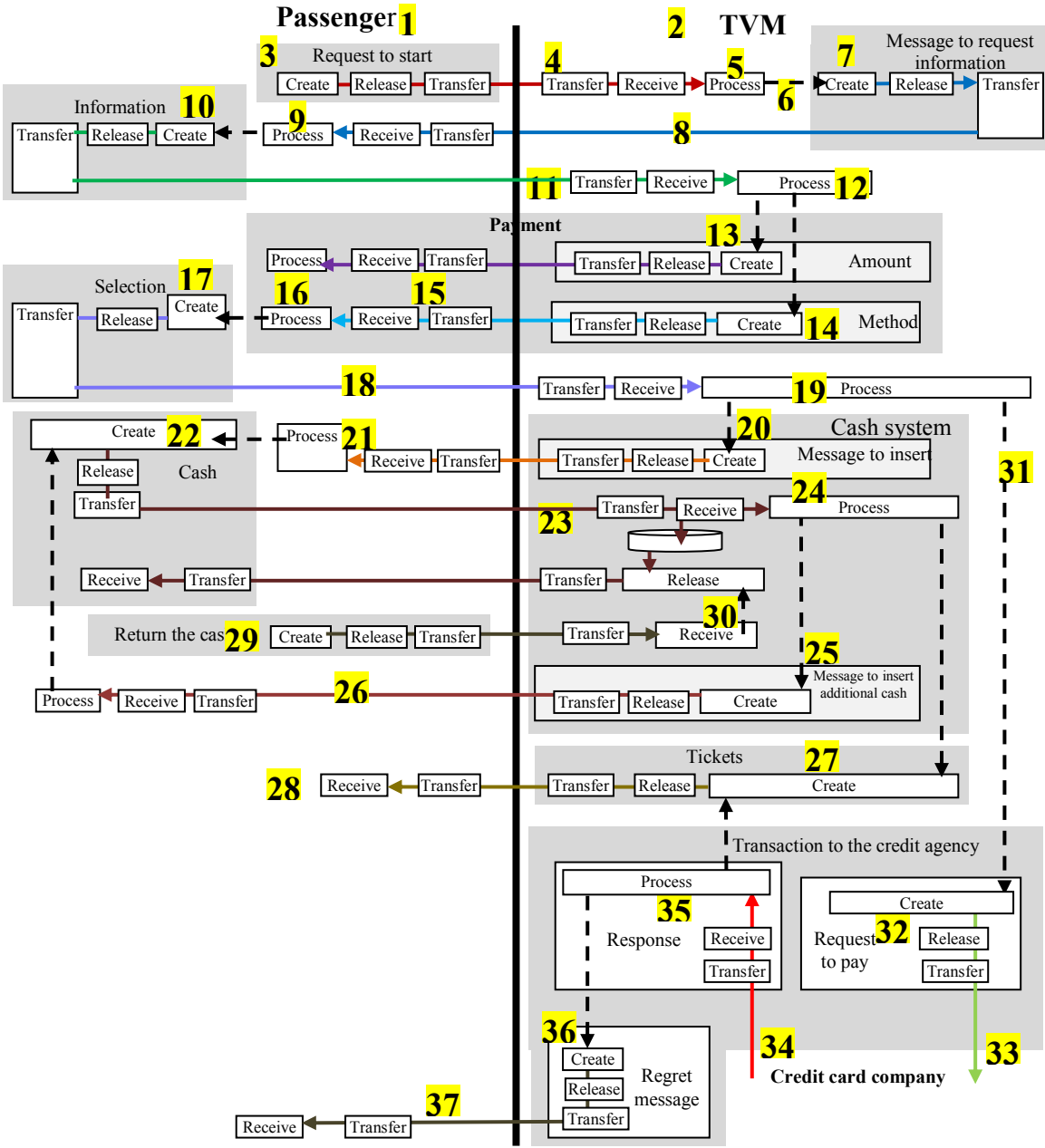

Fig. 3. FM representation of the TVM.

- If the selected payment method is credit card, then this triggers (31) the creation of a request for payment (32) that is sent to the credit card company (33). The TVM waits for a response (34); when it comes, the TVM processes (35) it as follows:

    (a) If the response to the request for payment is positive, then the TVM creates the tickets (27) and sends them to the passenger (28).

    (b) If the response to the request for payment is negative, then the TVM creates (36) a payment declined message and sends it to the passenger (37).

### D. Behavior Description

Note that Fig. 3 shows a *static schema* that does not embed dynamic behavior. It is a frame that constitutes the region in which events occur, "a possibility of fact—it is not the fact itself" [33]—in which a certain event is mapped to a subdiagram of the network of machines.

Behavior description is defined as the entire set of events that a system can perform and the order in which such events can be executed [34]. In system modeling, with FM methodology, behavior is modeled in a phase that occurs after the structural description is complete (e.g., Fig. 3) and involves modeling the "events space." Here, *behavior* involves the behavior of things during *events* when the system framework shown in Fig. 3 is acted upon. The chronology of events can be identified by orchestrating the sequence of events in their interacting processes.

In FM, an *event* is a *thing* that can be created, processed, released, transferred, and received. A *thing* becomes active in events. An event is specified by (1) its spatial area or subgraph, (2) its time, (3) the event's own stages, and (4) other possible qualities, e.g., intensity. For example, Fig. 4 shows the event of a passenger starting a transaction. Note that the region of the event is a subdiagram of Fig. 3. Note also that this event is not itself an elementary event because it is constituted of elementary events such as Create and Release.

### E. Control

Accordingly, the entire static representation of Fig. 3 is "event-ized", and the resulting events are utilized to control and manage the system.

For example, to save space, only *selection to pay in cash* of Fig. 3 is event-ized in Fig. 5, with the following events included:

- Event 1 ($V_1$): The TVM displays the instruction to insert cash.
- Event 2 ($V_2$): The passenger inserts cash that is received by the TVM.
- Event 3 ($V_3$): The TVM processes the cash.
- Event 4 ($V_4$): The TVM displays the instruction to insert more cash.

- Event 5 ($V_5$): The TVM creates tickets and sends them to the passenger.
- Event 6 ($V_6$): The passenger sends a request to cancel, hence to withdraw the cash.
- Event 7 ($V_7$): The TVM returns the cash to the passenger.

Accordingly, control of the chronology of the seven events can be developed as shown in Fig. 6. Going from left to right according to the flow of time,

- $V_1$, $V_2$, and $V_3$ (circles 1, 2, and 3) occur in sequence.
- This sequence is followed by *either* (circle 4) $V_4$ or $V_5$ (circles 5 and 6).
  - If $V_5$, then this is the end of the transaction.
  - If $V_4$, then it triggers (7) the creation of a repetition event (8), i.e., repeating $V_2$ and $V_3$. Note that this event has the attribute of possibility (9), that is, it may never occur.

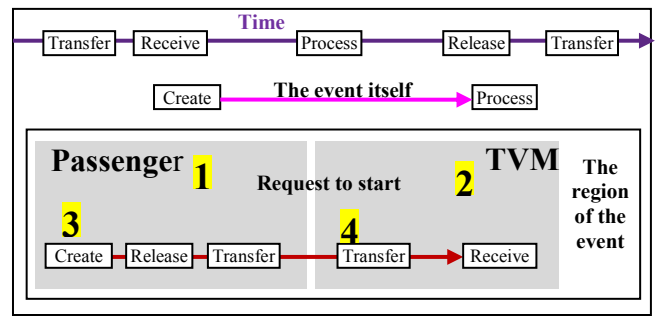

Fig. 4. Event of the passenger starting a transaction.

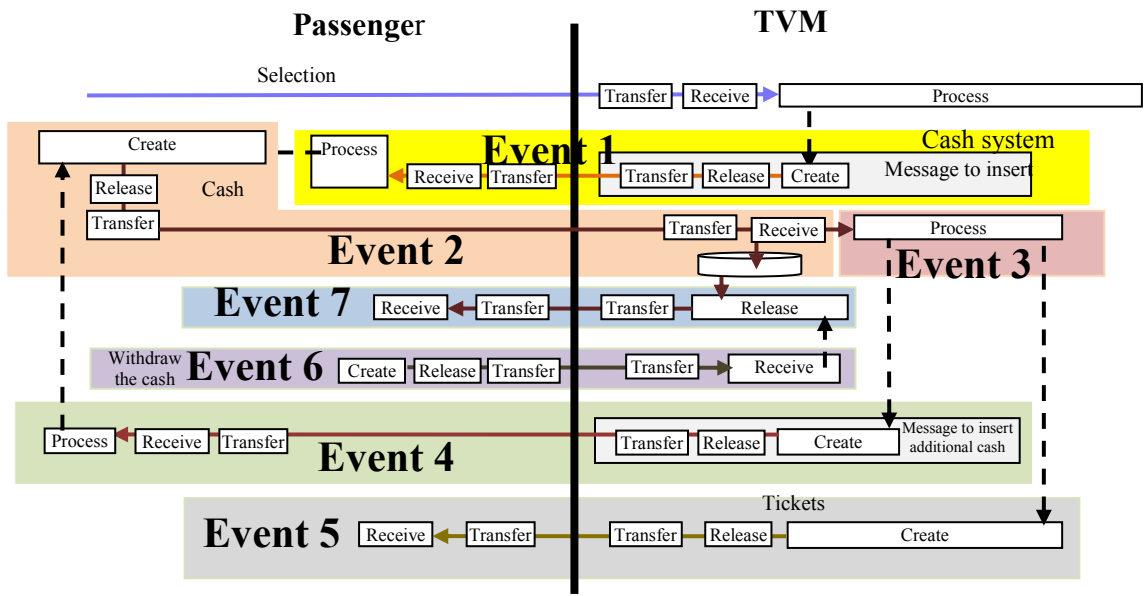

Fig. 5. Some events in the FM representation of the TVM.

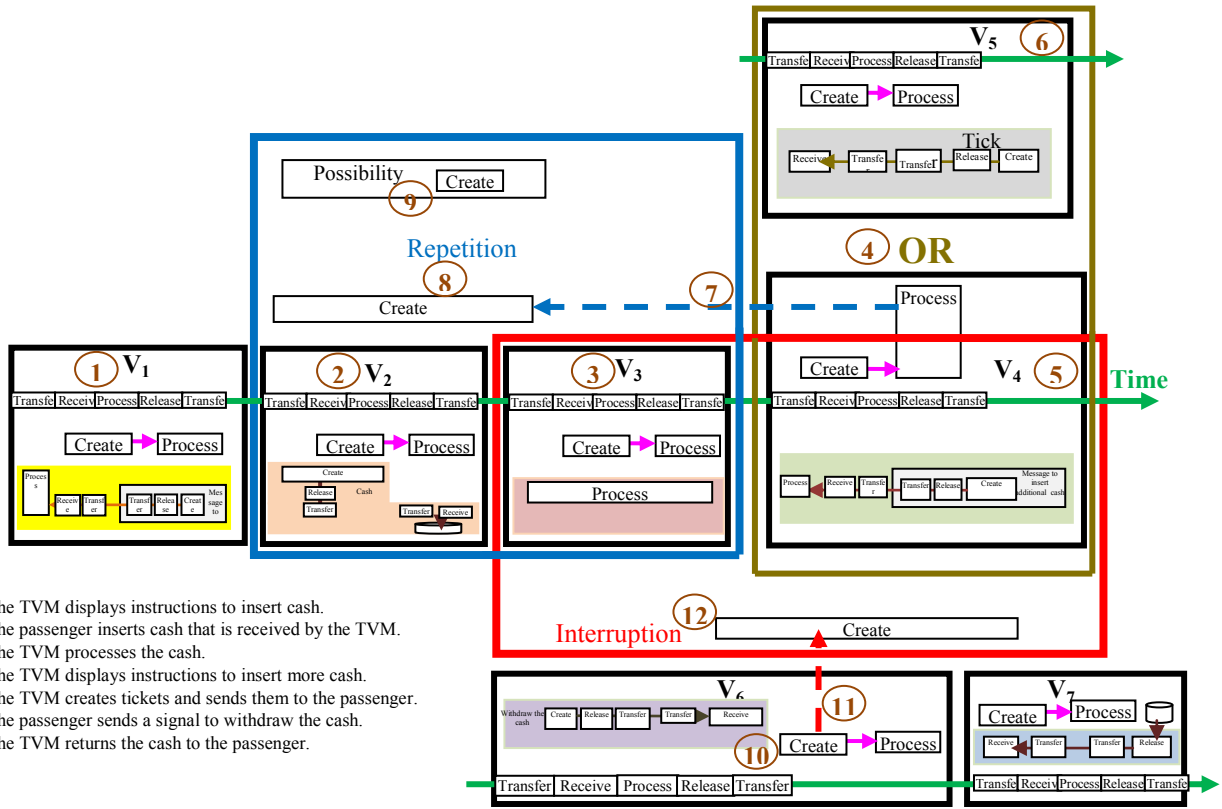

- $V_1$: The TVM displays instructions to insert cash.
- $V_2$: The passenger inserts cash that is received by the TVM.
- $V_3$: The TVM processes the cash.
- $V_4$: The TVM displays instructions to insert more cash.
- $V_5$: The TVM creates tickets and sends them to the passenger.
- $V_6$: The passenger sends a signal to withdraw the cash.
- $V_7$: The TVM returns the cash to the passenger.

Fig. 6. FM representation of the control of the chronology of events.

- When the cash is received, $V_6$ (in parallel with $V_3$ AND ($V_4$ OR V5)) is activated (10); thus, when the passenger signals to withdraw the cash, $V_6$ triggers (11) the interruption (12) of whatever flow machine ($V_3$, $V_4$, or $V_5$) is being executed at that moment, followed by $V_7$.

Fig. 7 simplifies this control specification by using an extension (e.g., adding a triggering interruption) of the classical specification of a chronology of events. Fig. 8 shows an FM simplification of Fig. 6.

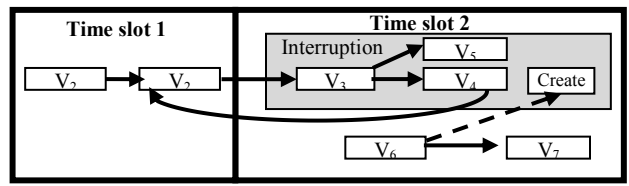

Fig. 7. Classical methods of representing the chronology of events.

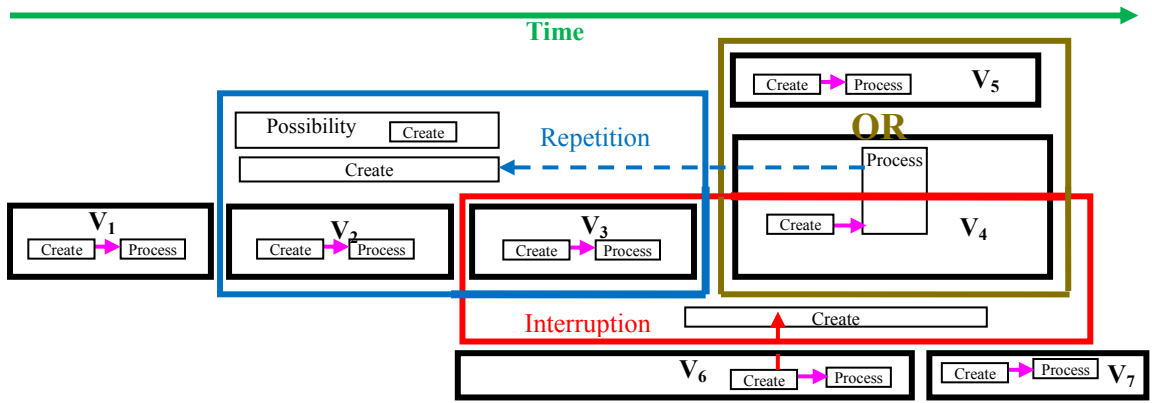

Fig. 8. FM simplification of the representation of control of a chronology of events.

### F.  General Comments

Note that FM modeling encompasses cross-world spheres of physical, digital, and social domains (cf. [4]). It can be used in the so-called cyber-physical system to orchestrate computers and physical systems to control physical processes that include feedback in both directions. It can provide a common model and methods for mechanical, environmental, civil, electrical, chemical, and industrial engineering.

Note that that, in general, the control module of a physical system is formed from physical things; e.g., wires carry basic measurement signals, and network components transfer messages between controllers, ports or terminals, and sensors. This also includes computer systems, message broadcasting, and services with request and reply messages.

### III.  Case Study: Electrical Power Plant

A model can be developed to serve many purposes, e.g., prediction and design. In our case, since the system to be modeled already exists, our purpose is to produce a conceptual representation of the system and its macroscopic behavior that can be used for many purposes, such as (physical and informational) control, maintenance, monitoring, management, and communication. The conceptual model can also help process engineering teams investigating a plant operational crisis, e.g., mysterious pipe vibration issues or problematic pieces of equipment or sections, by using it to simulate operational scenarios and for decision-making.

The system is an electricity-generating plant called Shuaiba South Power and Water Production Station (SSPWPS). The total compound electrical power generated by the plant can reach 804 MW (more details in [35]). An engineering schema of the modeled portion is shown in Fig. 9.

### A.  First-Level Model

The process of electrical power generation is modeled in Fig. 10, which shows fresh water flowing (circle 1) from the distillation station (not shown in this diagram) to two destination water tanks (2A and 2B).

- The water flows from the two-tank system (2A) to a *pipes/valves assembly* (3), then to a *common header valve system* (4). The *pipes/valves assembly* is a complex of pipes and valves used to control the rate of flow through several pipes. The *common header valve system* is used to unite flows from different sources. Thus, if there is only a single inflow, then other inflows in the figure would not be shown. Hereafter, to simplify the diagram, the interior structure of the *pipes/valves assemblies* and *common header valve system* will not be shown.

- The water also flows to the 2B water tank (6) through the pipes/valves assembly (7), then to the common header valve system (4).

Accordingly, the water in the common header valve system flows (5) to pumps (7) to increase flow pressure to reach another pipes/valves assembly (8), then another common header valve system (9). Simultaneously, the water in water tank 2B (6) flows through pumps (6A) then to a pipes/valves assembly (6B) to join other water in the common header valve system (9).

The mixed water in the common header valve system flows to the following:

(i) The demineralization plant (10)

(ii) The intake expansion tank (11), used to cool down the turbine (36)

(iii) The station water tank (12), where it is stored for firefighting purposes.

The water reaches another pipes/valves assembly (13) inside the demineralization (DM) plant, where it branches, as follows:

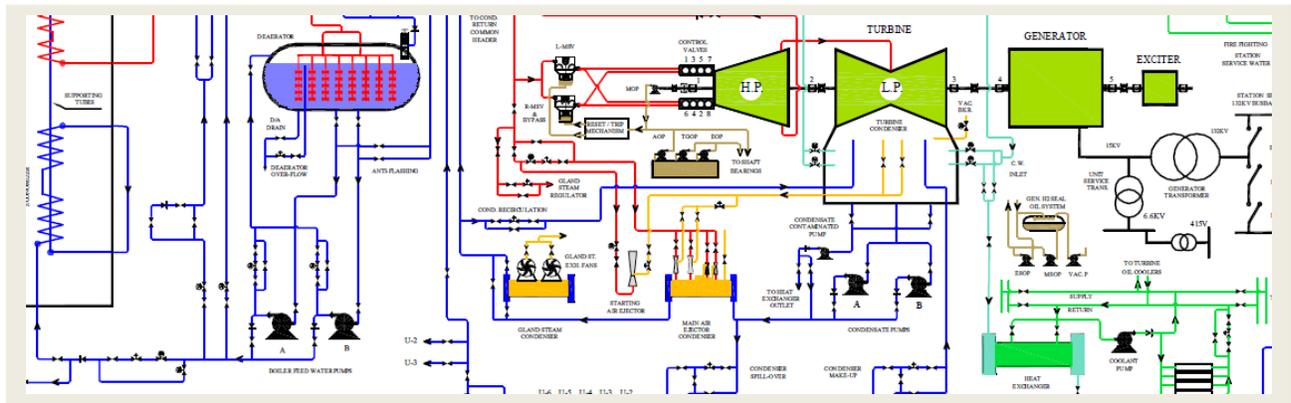

Fig. 9. Partial engineering schemata of the system to be modeled with FM.

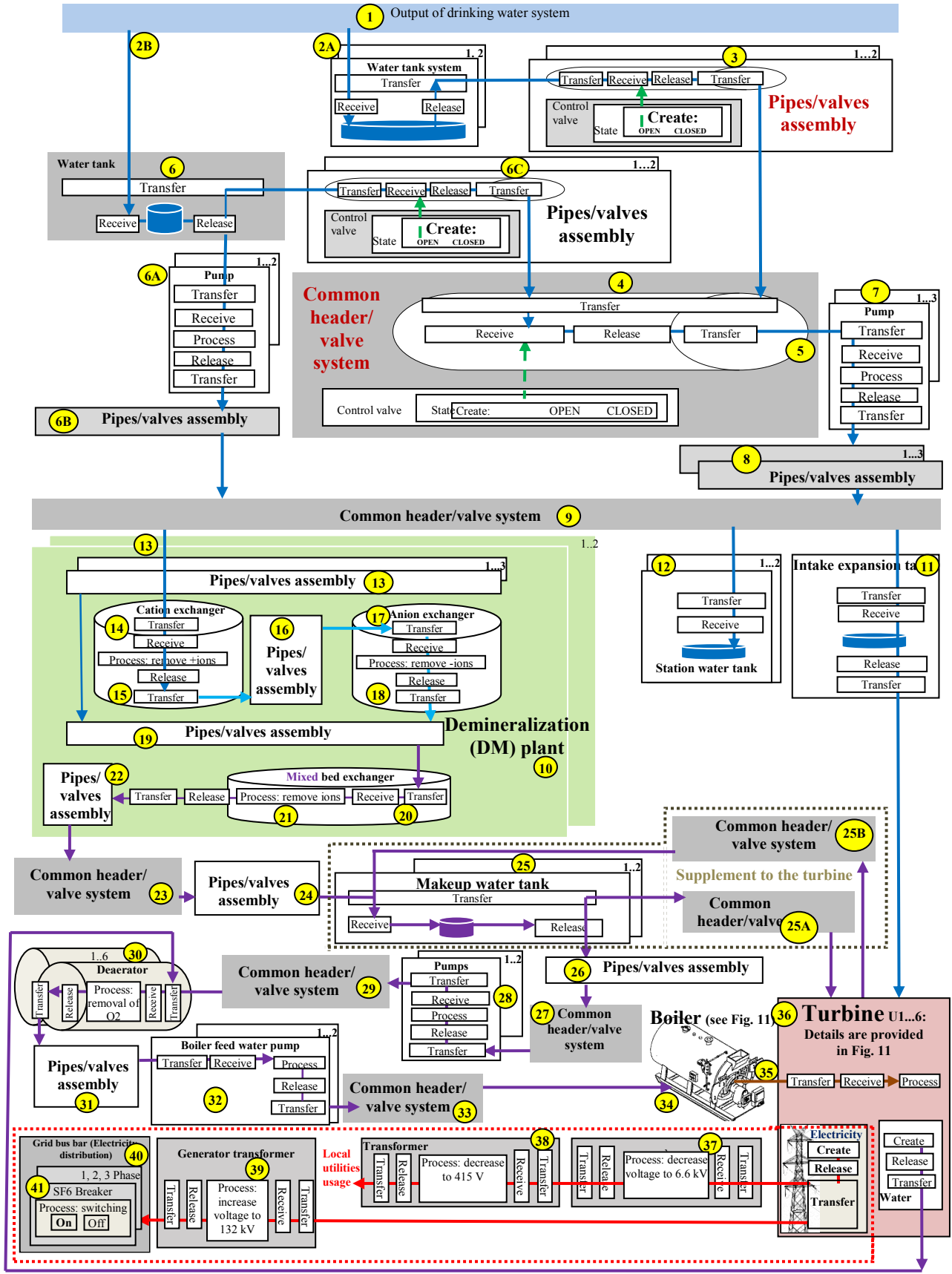

Fig. 10. FM representation of the electrical power plant system.

- The water flows to the cation exchanger (14), where the positive ions (cations) are removed (15), after which the water flows to a pipes/valves assembly (16) to reach the anion exchanger (17), where negative ions (anions) are removed (18).
- The resulting deionized water then flows through another pipes/valves assembly inside the DM plant (19) before reaching the mixed bed exchanger (20), where *both* cations and anions are removed (21). The deionized water then reaches the pipes/valves assembly (22).

*1) Flow of demineralized water:* From (22), the water flows through a common header valve system (23), then through another pipes/valves assembly (24) to reach the two makeup water tanks (25). These are used to supplement water needed for the turbine and boiler. Accordingly, the water flows in two directions:

- Supplement component to the turbine
- Supplement component to the boiler

*2) Supplement to the turbine:* The water flows from the makeup water tank (25) through a common header valve system (25A), then to the turbine (36). The turbine diverts excess water to a common header valve system (25B) that then returns it to the makeup water tank (25).

*3) Supplement to the boiler:* The water flows from the makeup water tank (25) through a pipes/valves assembly (26) and then to a common header valve system (27) connected to two water pumps (28). From the pumps, the water flows through a common header valve system (29) to the deaerators (30), which remove excess oxygen molecules from the water, removing the bubbles.

Water from the deaerators flows through a pipes/valves assembly (31) to the two boiler feed water pumps (32). The water then flows through a common header valve system (33), arriving at the boiler (34; a detailed FM subdiagram of the boiler will be shown), which produces high-pressure steam (35) that flows to the turbine (36) to lose its energy in running the turbine and converts to water that flows back to the deaerator (30).

Additionally, the high-pressure steam used to generate electricity in the turbine is then processed through a unit stepdown transformer (37) to reduce the voltage from 15 kV to 6.6 kV and send the electricity to another stepdown transformer (38) to further reduce the voltage to 415 V for local utilities usage.

Finally, the electricity generated by the turbine flows to another unit step-up transformer (39) to increase the voltage from 15 kV to 132 kV to be transported to the grid bus bar (40), passing by three circuit breakers (41).

To show that this modeling process can be applied to any level of description using the same technique, the turbine (34-35) will now be described in its own diagram.

*B. The Turbine*

Fig. 11 shows an FM representation of the turbine. It can be explained as follows:

*1) Heating the water:* The water from the common header (1; 33 in Fig. 10) reaches the first part of the turbine, the economizer (2). The economizer's function is to reduce the amount of energy needed to convert the water to steam, as follows:

- The water is heated (3) using the surrounding heat generated from the operation of the furnace (4) in order to convert it completely to steam with less fuel than would be necessary with low-temperature water.
- The furnace (4) is connected to 9 burners (5) that are fueled with gas (6). The ignition gun (7) is used to create a spark (8) to ignite the burners (5). Note that the heat generated by the furnace flows to the economizer (3).

In addition, the furnace receives atmospheric air (9). The air (top left of diagram) is sucked from the atmosphere by a forced draft fan (10) in the Boiler Air fuel gas system (11) to flow to the Air Preheater (12). The heated air flows to the damper (13) then (14) to the furnace, keeping the flame burning in the furnace (5). As a result, the burner produces exhaust gases (15) that flow to the damper (16), then to the air preheater (17), which heats the inlet air. The exhaust gases then flow to the chimney (18) to be released to the atmosphere (19).

*Creating steam:*

The high-temperature water in the economizer (2) flows to a boiler drum (20), where it is converted by heat (21) to wet steam (22) that flows to the primary super heater (23). This steam (24) flows though the attemperator (25) to reach the secondary super heater (26). The attemperator controls the temperature of the steam with water received from the attemperator spray water valve (27) originating from the common header (1).

*3.3 Further consideration*

Diagrams such as Figs. 10 and 11 can be applied in many areas, with the simplest being documentation, where "Documents are a means to present information instead of being containers of information" [7]. However, here we suggest that the FM diagram is an important tool for maintenance/operations and management.

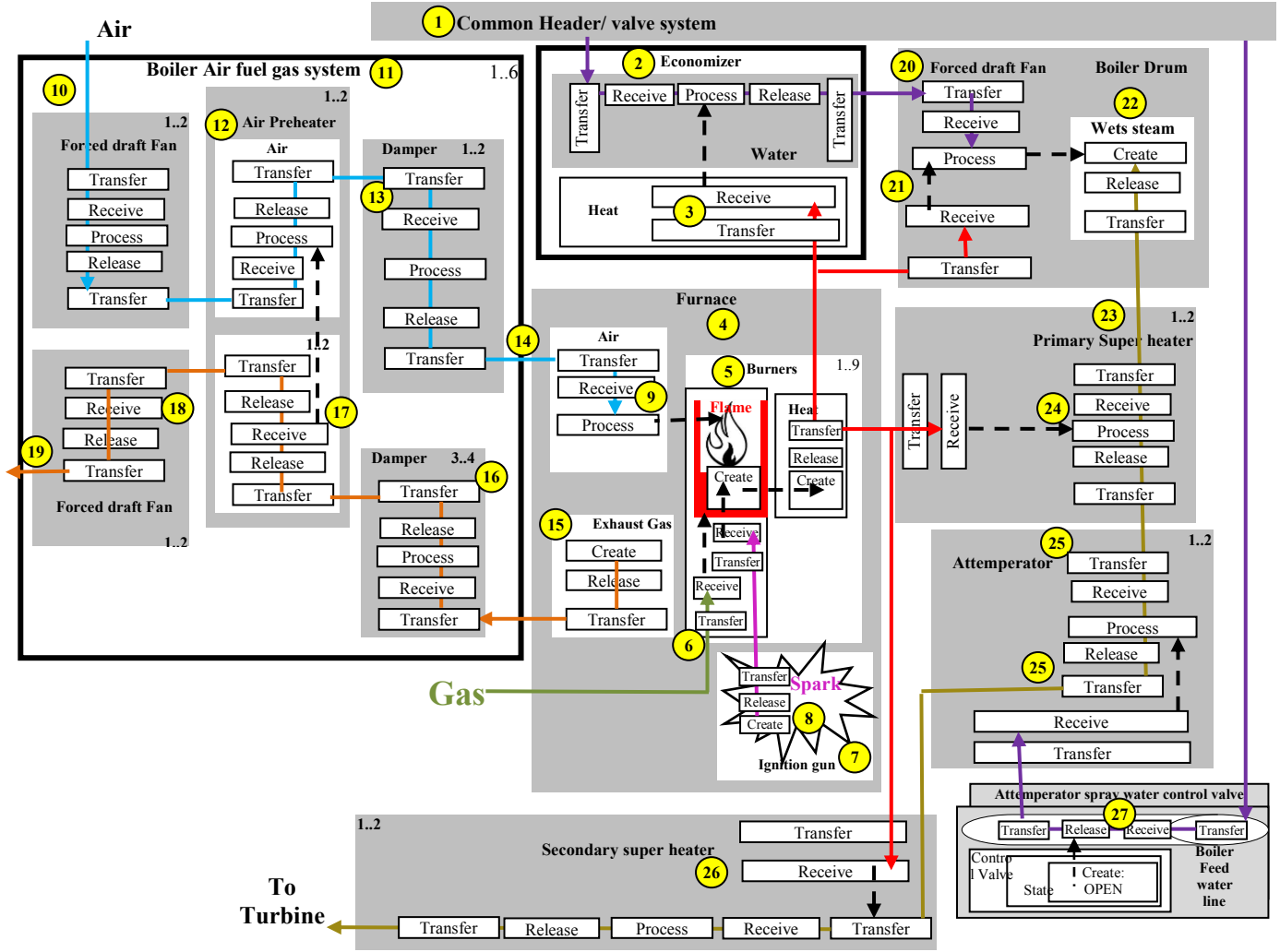

Fig. 11. FM representation of the turbine.

The flows shown in Figs. 10 and 11 can be event-ized as discussed in Section II, according to meaningful events, in order to impose control over different components of the system. Because of space limitation, we focus here on a sample case: the situation of keeping track of parts replacement over time. Equipment needs to be regularly maintained or replaced, and equipment history is a major issue for situations such as scheduled maintenance. In addition, from a conceptual point of view, difficulties arise "When equipment is scheduled for maintenance, it is looked at on an individual basis without evaluating its impact on a system" [36].

There is a need for holistic views and systems thinking in the planning of service and maintenance activities… more efforts are desired to support the development in this direction and to quantify the benefits of being more holistic and flow-oriented [in] the planning of service and maintenance activities. [37].

Clearly, the FM approach with its holistic representation of systems can help with this type of problem. In this section, we briefly demonstrate how FM can be used to conceptualize the situation of "changing parts" over time. According to Tommila and Alanen [7],

Elements of a system may be changed over time without the system losing its identity. Therefore, the elements of a system can be understood as place holders for actual component individuals that, in many cases, are instantiations of a commercial product or device type and have a manufacturer's serial number. For example, [Fig. 12] shows a functional pump object P101.

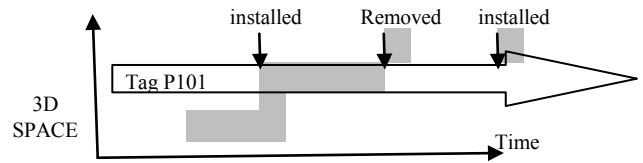

Fig. 12. Replacing a pump (redrawn, partial from [6]).

It is distinct from the individual "pump 1" that was first installed as P101 and later replaced by a spare part item "pump 2". Including the time dimension seems useful not only for design and system modeling but also for configuration management and traceability during system operation.

Let us assume that Tommila and Alanen's [7] pump object P101 is one of the two pumps shown at (28) in Fig. 10, shown again in Fig. 13. Fig. 14 shows the FM representation of the history of replacing pump object P101 over time. Note that the same FM notations are used to represent this history.

In the figure, the sphere of pump P101 (circle 1) includes the pump machine itself (2) and the water machine (3) as part of the description of the total system. Event 1 (3) is a "happening" that occurs to that pump during a certain period of time beginning at (5) and ending at (6). The event involves receiving the pump (7) and installing it (8).

This event is followed at a later period of time by event 2, which comprises removal of the pump (9). Note that the occurrence of this sequence of events is represented perpendicularly over the static description of Fig. 10, as reflected by the downward right-angled arrows connecting the flow of time. Similarly, during a later period of time, events occur until event *n*.

Each event can include additional information such as who performs the work and name of the maintenance contractor. Thus, the FM diagram "grows" vertically to represent time and to register changes of different parts in the system description. The result is a clear conceptual and orderly foundation of the operations of the system and its changes over time. Of course this foundation would have to be translated into a practical informational and control scheme.

## IV. CONCLUSION

The FM model can be utilized uniformly to describe physical engineering systems and their behavior for purposes of integrating maintenance and operations into a total system that comprises humans, physical objects, and information. Conceptual complexity is resolved through simple, uniform notations applied across macro- and micro-levels of detail. FM diagrams become more complex as specifications become more complete. It is possible to utilize granularity levels, refinement, and zooming to reduce the appearance of complexity.

Still, a great deal of work is needed to apply the FM approach in practical situations. Nevertheless, the FM model seems promising and merits further development in diverse engineering applications.

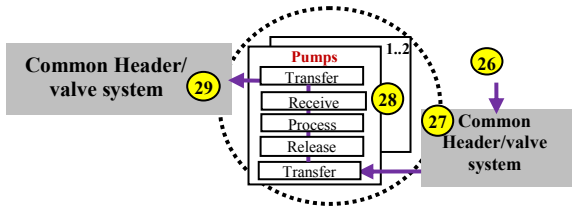

Fig. 13. Portion of Fig. 10 that includes the replaced pump.

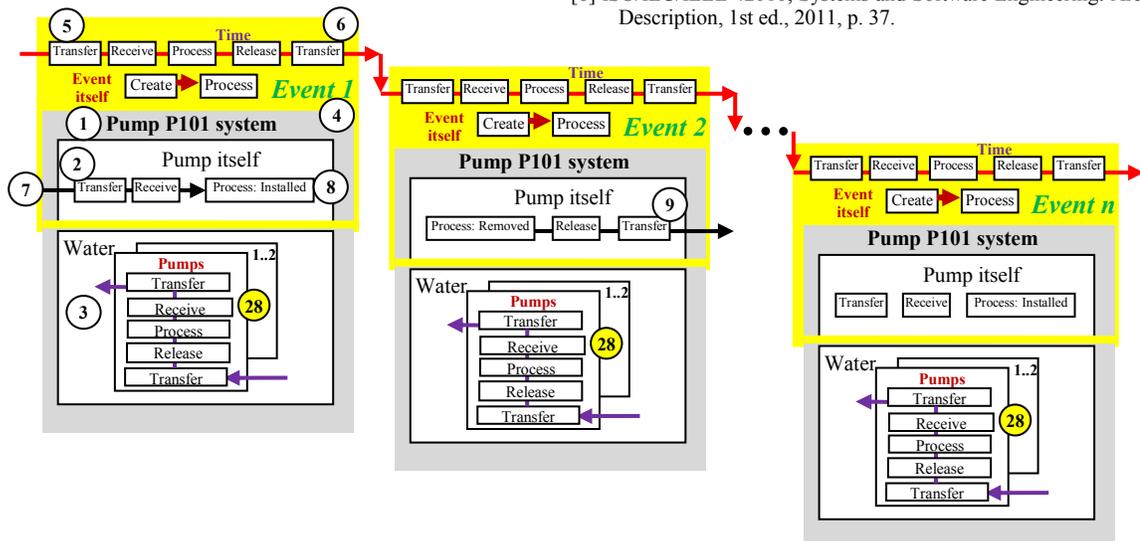

Fig. 14. FM representation of the events of replacing the pump 1 P101.

## AUTHORS PROFILE

Sabah Al-Fedaghi holds an MS and a PhD in computer science from the Department of Electrical Engineering and Computer Science, Northwestern University, Evanston, Illinois, and a BS in Engineering Science from Arizona State University, Tempe. He has published two books and more than 270 papers in journals and conferences on software engineering, database systems, information systems, computer/ information privacy, security and assurance, information warfare, and conceptual modeling. He is an associate professor in the Computer Engineering Department, Kuwait University. He previously worked as a programmer at the Kuwait Oil Company and headed the Electrical and Computer Engineering Department (1991–1994) and the Computer Engineering Department (2000–2007).

Abdulaziz Alqallaf holds Bachelor's and Master's in computer engineering from the Department of Computer Engineering, Kuwait University. He has been working since 2015 as a computer engineer in the Instrument Maintenance Department, Ministry of Electricity and Water, Kuwait. His interests include computer networks, security and software engineering.